\documentclass[pdflatex,sn-mathphys-num]{sn-jnl}

\usepackage{graphicx}%
\usepackage{multirow}%
\usepackage{amsmath,amssymb,amsfonts}%
\usepackage{amsthm}%
\usepackage{mathrsfs}%
\usepackage[title]{appendix}%
\usepackage{xcolor}%
\usepackage{textcomp}%
\usepackage{manyfoot}%
\usepackage{booktabs}%
\usepackage{algorithm}%
\usepackage{algorithmicx}%
\usepackage{algpseudocode}%
\usepackage{listings}%

\begin{document}

\title[Article Title]{Andreev bound states in a superconducting qubit at odd parity}

\author[1]{\fnm{Manuel} \sur{Houzet}}

\author[1]{\fnm{Julia~S.} \sur{Meyer}}

\author[2]{\fnm{Yuli~V.} \sur{Nazarov}}

\affil[1]{\orgdiv{Univ.~Grenoble Alpes, CEA, Grenoble INP, IRIG, PHELIQS}, \orgaddress{\city{F-38000 Grenoble}, \country{France}}}

\affil[2]{\orgdiv{Kavli Institute of Nanoscience}, \orgname{Delft University of Technology}, \orgaddress{\city{2628 CJ Delft}, \country{Netherlands}}}

\abstract{
The quantum mechanics of the Josephson effect is the core ingredient for quantum technologies
with superconducting circuits. A new avenue was recently opened in this field by predicting that the
Josephson quantum mechanics in the odd parity sector, when a quasiparticle in trapped in an
Andreev bound state, is fundamentally different from the conventional one in the even sector. The focus was then on a Josephson junction surrounded by an electromagnetic environment formed of a collection of bosonic modes, including the case of an ohmic environment. Here we consider the distinct case of a superconducting qubit made of a single Josephson junction whose environment reduces to a capacitance. We find a novel structure for the low-lying discrete states in the odd sector, which is altogether different from the one that appears in the even sector. Our study of the bound-state spectrum ranges from the Coulomb-dominated (Cooper pair box) to the Josephson-dominated (transmon) regime. Our prediction could be tested in forthcoming experiments with superconductor/semiconductor/superconductor junctions, which have been studied intensively in recent years, both using nanowires as well as two-dimensional electron gases.
}

\keywords{Josephson effect, superconducting qubit, Andreev bound state, quasiparticle poisoning}

\maketitle

\section{Introduction}
\label{sec:1}

The simplest design of a superconducting qubit is a Josephson junction that separates two superconducting islands. In that device, the competition between charge localization due to Coulomb repulsion, characterized by a charging energy $E_C$, and charge delocalization due to Cooper pair tunneling, characterized by the Josephson energy $E_J^*$, realizes a quantum-mechanical problem that yields the formation of discrete states. The two lowest states form the qubit. At low temperature, a large superconducting gap $\Delta$ protects the device from dissipation due to quasiparticles. In the so-called ``Cooper-pair box'', corresponding to $E_J^*\ll E_C$~\cite{Bouchiat1998,Nakamura1999}, the discrete levels are highly sensitive to fluctuations of an external charge in the environment of the junction, which is detrimental for the quantum coherence of the qubit. In a ``transmon'', corresponding to $E_J^*\gg E_C$~\cite{Koch2007,Paik2011}, the charge sensitivity is exponentially suppressed, which results in much better coherence properties. By now transmons have become an essential building block of many quantum technologies with superconducting circuits.

Quasiparticle poisoning in superconducting qubits has been mostly discussed in relation with the experimental observation of a much larger quasiparticle concentration than expected at low temperature~\cite{Day2003}. As a result, the qubit lives in two different parity sectors, depending whether an even or odd number of those quasiparticles were transferred across the junction. A signature of these parity sectors is observed in the charge dispersion in the qubit excitation spectrum: it is $2e$-periodic in each parity sector, while the dispersions between even and odd sectors differ by an $e$-shift, where $e$ is the elementary charge~\cite{Aumentado2004,Schreier2008}. Furthermore, each quasiparticle transfer across the junction may induce energy relaxation of the qubit~\cite{Martinis2009,Catelani2011,Catelani2011b,Lenander2011}. Various mitigation strategies are currently being investigated to suppress the detrimental effects of parity switching events on qubit coherence properties, see e.g.~\cite{McEwen2024}. 

An additional ingredient in the physics of Josephson junctions is the formation of fermionic Andreev bound states~\cite{Beenakker1991,Furusaki1991} localized at the junction. This ingredient introduces a distinct notion of  parity compared with the one discussed above. Indeed, instead of being delocalized in an island, an excess quasiparticle may get trapped in a bound state. A many-body state whereby a single quasiparticle is trapped in the lowest Andreev state (odd-parity ground state) is long lived due to parity conservation in superconductors~\cite{Averin1992}, despite having an energy larger by $\sim\Delta$ compared with the even-parity ground state. The Andreev level dispersion with a classical superconducting phase difference $\varphi$ applied to the junction has been much studied. In particular, in the short-junction limit relevant for tunnel junctions, these bound states merge with the continuum of excitations above $\Delta$ at $\varphi=0$. This is not the case anymore when one takes into account quantum fluctuations of the phase induced by the electromagnetic environment of the junction. So far this question was addressed in the case of a Cooper-pair box made of a single-channel junction and occupied with an odd number of elementary charges~\cite{Pesin2004}, where it was shown that a discrete state detaches from the continuum of excitations at any value of the charge gate. They found that the charge dispersion of the bound state dispersion is $e$-periodic with the gate and that its binding energy presents a cusp at the charge degeneracy points. Furthermore, our previous work on the Josephson quantum mechanics at odd parity~\cite{Houzet2024} predicted an intriguing structure of Andreev bound states when a Josephson junction is galvanically coupled to an electromagnetic environment formed of a collection of harmonic oscillators, including the case of an ohmic environment.

In this work we generalize the results of Ref.~\cite{Pesin2004} by extending our theory~\cite{Houzet2024} to the case of a capacitive coupling to the electromagnetic environment. Our analysis applies to a superconducting qubit with arbitrary ratio $E_J^*/E_C$ and number of channels. In particular we find that several bound states may appear in a given channel when the contribution of that channel dominates over the other ones and $E_J^*$ becomes comparable or larger that $E_C$. Furthermore, the structure of these bound states is strikingly different from the one determined from the Mathieu equation applicable in the even sector~\cite{Koch2007}. These multiple bound states could be revealed by microwave spectroscopy. Due to their $e$-periodic dispersion with gate charge, poisoning related with excess quasiparticles in the islands would not allow distinguishing two sectors as in Ref.~\cite{Schreier2008}. 

In the rest of the article, we derive an effective Hamiltonian that allows determining the bound states in a superconducting qubit at odd parity in Section~\ref{sec:2}, we analyze the binding energy in various regimes, both analytically and numerically, in Section~\ref{sec:3}, and we present our conclusions and perspectives in Section~\ref{sec:4}.

\section{Hamiltonian}
\label{sec:2}

In this section we derive a low-energy effective eigenvalue problem that encodes the Andreev spectrum in the odd parity sector of a superconducting qubit made with a single Josephson junction, cf. Eqs.~\eqref{eq:main2} and \eqref{eq:main3} below. This eigenvalue problem greatly simplifies our study as it only depends on an electromagnetic degree of freedom, while the fermionic ones have been integrated out. Our derivation also highlights the differences with our recent study on Josephson quantum mechanics at odd parity~\cite{Houzet2024} due to the different nature of the coupling to the electromagnetic environment (capacitive here, galvanic there). Namely in the present case the environment Hamiltonian depends on where the additional quasiparticles resides.

We first consider the case where one side of the Josephson junction is a Coulomb island with total capacitance $C$. Its electrostatic energy is
\begin{equation}
\hat H_C=\frac1{2C}(\hat Q-Q_g)^2,
\end{equation}
where $\hat Q$ is the operator of the island charge and $Q_g$ is an external gate charge.

In a normal island, each electron state is indexed by orbital label $\ell$ and spin $\sigma=\uparrow,\downarrow$, and has energy $\xi_\ell $ (measured from the Fermi level). Each $\ell $ spans a Nambu box with four states: the vacuum state $|0\rangle_\ell $, the doubly occupied state $|2\rangle_\ell  = a^{\dagger}_{\ell \uparrow} a^{\dagger}_{\ell  \downarrow}|0\rangle_\ell $, and two doublet states $|\sigma\rangle_\ell  = a^{\dagger}_{\ell  \sigma}|0\rangle_\ell $. Here $a^{\dagger}_{\ell \sigma}$ is an electron creation operator for the state labelled with $\ell $ and $\sigma$. In the presence of a superconducting pairing gap $\Delta$, the states $|0\rangle_\ell $ and $|2\rangle_\ell $ hybridize. Following BCS theory, we introduce the family of degenerate wave functions $|\varphi\rangle$:
\begin{equation}
|\varphi\rangle = \prod_\ell  |g\rangle_\ell \qquad \mathrm{with}\qquad |g\rangle_\ell =u_\ell  |0\rangle_\ell  + v_\ell  e^{i\varphi} |2\rangle_\ell ,
\end{equation}
where $u_\ell ,v_\ell  = \sqrt{(1\pm\xi_\ell /E_\ell )/2}$ are coherence factors and $E_\ell =\sqrt{\xi_\ell ^2+\Delta^2}$ is an excitation energy. The operator of the island charge is then represented as
\begin{equation}
\label{eq:Qe}
\hat{Q} = 2e i \partial_{\varphi}.
\end{equation}
Its eigenvalues are {multiples of $2e$}  provided that the $2\pi$-periodicity in superconducting phase variable $\varphi$ is imposed. In this approach, a single quasiparticle state in the Nambu box $\ell $ reads
\begin{equation}
|\ell \sigma\rangle = 
|\sigma\rangle_\ell \prod_{\ell'\neq\ell} |g\rangle_{\ell'}.
\end{equation}
For us it is important to note that the charge operator for this family of states reads
\begin{equation}
\label{eq:Qo}
\hat{Q} = 2e i \partial_{\varphi} -e.
\end{equation}
That is, a quasiparticle bears one electron charge. We introduce the usual quasiparticle creation operators (using $+$ for $\uparrow$ and $-$ for $\downarrow$),
\begin{subequations}
\label{eq:bg-qp}
\begin{align}
\label{eq:qp-ep}
\hat \alpha^\dagger_{\ell \sigma} = u_\ell  \hat a^{\dagger}_{\ell \sigma} - \sigma v_\ell  e^{-i\varphi} \hat a_{\ell -\sigma}, 
\end{align}
\end{subequations}
to check explicitly that $\hat \alpha_{\ell \sigma}|\varphi\rangle =0$ and $\hat \alpha^\dagger_{\ell \sigma}|\varphi\rangle= |\ell \sigma\rangle$.

\begin{figure}
\centering
\includegraphics[width=.7\columnwidth]{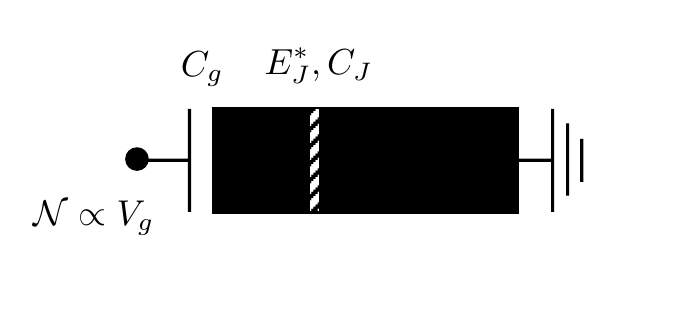}
\caption{\label{F0}
The circuit consists of a superconducting island contacted to a superconducting lead through a tunnel junction (hatched region) characterized by Josephson energy $E_J^*$ and capacitance $C_J$. The external gate charge ${\cal N}=-C_gV_g/e$ is controlled by the voltage $V_g$ applied to an electrostatic gate with capacitance $C_g$. The Coulomb energy of the island is $E_C=e^2/2C$ with $C=C_J+C_g$.    
}
\end{figure}

Let the island be on the {left} side of the Josephson junction, see Fig.~\ref{F0}. As to the {right} side, we connect an open lead with fixed phase $\varphi=0$. The relations for the electron ($\hat c^\dagger_{m\sigma}$) or quasiparticle ($\hat \gamma^\dagger_{m\sigma}$) creation operators in a state with orbital label $m$, spin $\sigma$, and energy $\xi_m$ in normal state or $E_m=\sqrt{\xi_m^2+\Delta^2}$ in superconducting state, are identical to Eqs.~\eqref{eq:bg-qp} at $\varphi=0$. (Here we assume that the island and the lead are made of the same material and have identical gap $\Delta$.) Now we can express the tunneling Hamiltonian for a single channel,
\begin{equation}
\hat H_T = \frac{t_0}{\sqrt{{\cal V}_L{\cal V}_R}} \sum_{\ell m\sigma} \left(\hat a^{\dagger}_{\ell \sigma} \hat c_{m\sigma} + \hat c^\dagger_{m\sigma}  \hat a_{\ell \sigma} \right),   
\end{equation}
where $t_0$ is the tunnel matrix element in a given channel of the junction and ${\cal V}_L,{\cal V}_R$ are normalization volumes of the island and the lead, in terms of quasiparticle operators, and separate it into two parts: $\hat H_T=\hat H_T^{(qp)} +\hat H_T^{(cp)}$ with
\begin{align}
\hat H_T^{(qp)} = \frac{t_0}{\sqrt{{\cal V}_L{\cal V}_R}} \sum_{\ell m\sigma}   \left(u_\ell  u_m-v_\ell  v_m e^{i\hat\varphi}\right)\hat \alpha^\dagger_{\ell \sigma} \hat \gamma_{m\sigma}+ \mathrm{H.c.},\\
\hat H_T^{(cp)} = \frac{t_0}{\sqrt{{\cal V}_L{\cal V}_R}} \sum_{\ell m\sigma} \sigma  \left(u_\ell  v_m+ v_\ell  u_m e^{i\hat\varphi}\right)\hat  \alpha^\dagger_{\ell \sigma} \hat \gamma^\dagger_{m-\sigma}+ \mathrm{H.c.}.
\end{align}
The first term describes quasiparticle tunneling, while the second term describes the creation and annihilation of pairs of quasiparticles on either side of the junction.

We restrict further considerations on the odd sector, where a single quasiparticle resides on either side of the junction, i.e., in the island or the lead. The Hamiltonian associated with the energy of that quasiparticle is
\begin{equation}
\label{eq:H0}
\hat H_0=\sum_{\ell \sigma}E_\ell |\ell \sigma\rangle\langle \ell \sigma|+\sum_{m\sigma}E_m|m\sigma\rangle\langle m\sigma|,
\end{equation}
depending on whether the quasiparticle resides in the island or in the lead.

The contribution of $\hat H_T^{(cp)}$ is taken in second-order approximation. Here the effect of the single quasiparticle is negligible.
If $\Delta$ is much bigger than the charging energy, $E_C=e^2/2C$, it reduces to the familiar Josephson coupling term,
\begin{equation}
\hat H_J = - E^*_J \cos\hat \varphi.
\end{equation}
While the quasiparticle resides in one particular channel, the Josephson energy $E_J^*$ is obtained by summing over all channels. Namely $E_J^*\geq E_J$, where $E_J=(\pi \nu t_0)^2\Delta$ is the Josephson energy for a single channel. Here $\nu$ is the normal density of state (per spin) in the island and the lead. We define $N_{\rm ch}=E^*_J/E_J$ henceforth as an effective number of channels of the junction. (While the identification of $N_{\rm ch}$ with a number of channels strictly holds only if all channels have the same transparency, the provided formulas with the introduced definition of $N_{\rm ch}$ are general.)\footnote{As a side note: the contribution of $\hat H_T^{(cp)}$ can be evaluated in the regime $\Delta \simeq E_C$ as well. In this case, the energy difference denominator contains the charging energy differences corresponding to addition/extraction of a charge to/from the island. This would modify the effective Josephson energy and lead to a more complex operator $\hat H_J$.}

As to $\hat H_T^{(qp)}$, it can be directly projected onto the targeted space, giving the tunneling amplitudes of the quasiparticle transfers in both directions. We incorporate a low-energy approximation $u_\ell ,v_\ell  \approx 1/\sqrt{2}$ (and $E_\ell \approx \Delta+\xi_\ell ^2/2\Delta$) to find 
\begin{equation}
\hat H_T^{\rm eff} =-\frac{t_0}{\sqrt{{\cal V}_L{\cal V}_R}} \sin\frac{\hat \varphi}2\sum_{\ell ,m,\sigma}ie^{i\hat \varphi/2} |\ell \sigma\rangle \langle m\sigma| + \mathrm{H.c.}. 
\end{equation}
The total effective Hamiltonian in the space of interest is thus
\begin{equation}
\label{eq:Heff}
\hat H_{\rm tot}=\hat H_0+\hat H_C+\hat H_J +\hat H_T^{\rm eff}
\end{equation}
with $\hat Q$ in $\hat H_C$ given by Eq.~\eqref{eq:Qe} or \eqref{eq:Qo}, depending whether the quasiparticle resides in the lead or island.

We may then look for an eigenfunction of Hamiltonian \eqref{eq:Heff} with fixed spin $\sigma$ and energy $E$ as 
\begin{equation}
|\Psi\rangle=\sum_\ell |\psi_\ell (\varphi)\rangle|\ell \sigma\rangle+\sum_m|\psi_m(\varphi)\rangle|m\sigma\rangle,
\end{equation}
where $|\psi_\ell (\varphi)\rangle,|\psi_m(\varphi)\rangle$ are wavefunctions in the phase $\varphi$ degree of freedom that are labelled by quasiparticle states. These quasiparticles states solve
\begin{subequations}
\begin{eqnarray}
\!\!\!\!\!\!\left[\Delta-E+\frac{\xi_\ell ^2}{2\Delta}+ \hat H_J+{E_C}( -2 i \partial_{\varphi}+1-{\cal N})^2\right]|\psi_\ell \rangle
&=&-\frac{ie^{-i \varphi/2}t_0\sin\frac{ \varphi}2}{2\sqrt{{\cal V}_L{\cal V}_R}} \sum_{m}  |\psi_m\rangle,\quad\\
\left[\Delta-E+\frac{\xi_m^2}{2\Delta}+\hat H_J+{E_C}(-2 i \partial_{\varphi}-{\cal N})^2\right]|\psi_m\rangle
&=&\frac{ie^{i \varphi/2}t_0\sin\frac{\varphi}2}{2\sqrt{{\cal V}_L{\cal V}_R}} \sum_{\ell }  |\psi_\ell \rangle,\quad
\end{eqnarray}
\end{subequations}
where  ${\cal N}=-Q_g/e$ is an effective gate voltage (in units of $e$).
We may then arrive at an eigenproblem defined on a spinor wavefunction in the island/lead space, 
\begin{equation}
|\Phi(\varphi)\rangle=\left(\begin{array}{c}{\cal V}_R^{-1/2}\sum_\ell |\psi_\ell (\varphi)\rangle\\{\cal V}_L^{-1/2} \sum_m|\psi_m(\varphi)\rangle\end{array}\right),
\end{equation} 
which does not depend on the quasiparticle degree of freedom anymore:
\begin{equation}
\label{eq:main}
\left(\sqrt{\Omega + \check H}-\sqrt{2E_J}\sin\frac{\hat\varphi}2\check s\right)|\Phi\rangle  =0.
\end{equation}
Here, $\Omega=\Delta-E+E_{\rm min}({\cal N})$ is the binding energy and
\begin{subequations}
\begin{eqnarray}
\check H &=& \hat H_J + E_C\left(-2 i \partial_{\varphi}+\check q-{\cal N}\right)^2-E_{\rm min}({\cal N}),\\
\check q&=&\frac12(1+\tau_z),\\
\check s&=&-\sin\frac{\hat\varphi}2\tau_x+\cos\frac{\hat\varphi}2\tau_y,
\end{eqnarray}
\end{subequations}
where $\check H,\check q,\check s$ are $2\times 2$ matrices in the island/lead space, in which we introduced Pauli matrices $\tau_{x,y,z}$, and the energy $E_{{\rm min}}({\cal N})$ is chosen to guarantee that the lowest eigenvalue of $\check H$ is 0. Let us note that the Hamiltonian formulation in Eq.~\eqref{eq:main} is explicitly $2\pi$-periodic and fully compatible with the periodic boundary condition
\begin{equation}
\label{eq:BC}
|\Phi(\varphi=0)\rangle = |\Phi(\varphi=2 \pi)\rangle.
\end{equation}
Equation~\eqref{eq:main} does not immediately compare with Eq.~(4) of Ref.~\cite{Houzet2024}, which was derived for a different electromagnetic environment made of a collection of harmonic oscillators {\it \`a la} Caldeira-Leggett.

To establish a link between them, let us make a unitary transformation that brings the charge operator $\check Q=\hat Q-e\check q$ to the standard form and changes the matrix $\check s$:
\begin{equation}
\check S^{-1} \check Q\check S= 2 ei \partial_{\varphi}
\quad {\rm and}\quad
\check S^{-1} \check s\check S =\tau_y
\quad {\rm with}\quad
\check S=\exp\left[-i{\varphi}\check q/2\right].
\end{equation}
Na\"ively it seems that we can diagonalize $\tau_y$ and solve the resulting Hamiltonian for opposite chiralities $s=\pm 1$, as in Ref.~\cite{Houzet2024}, in which chirality $s$ was introduced as a superposition index of quasiparticle states between the island and the lead. However, this is not so simple since the unitary transformation also modifies the boundary condition~\eqref{eq:BC} and is not diagonal in the space of chiralities:
\begin{equation}
\label{eq:BC2}
|\Phi(\varphi=2\pi)\rangle = - \tau_z|\Phi(\varphi=0)\rangle.
\end{equation}
Still, we notice that the basis vectors compatible with the new boundary conditions~\eqref{eq:BC2} read
\begin{equation}
\exp[i n \varphi]\begin{pmatrix} 0 \\ 1\end{pmatrix},\qquad \exp[i (n+1/2) \varphi]\begin{pmatrix} 1 \\ 0\end{pmatrix}
\end{equation}
($n$ integer). These basis vectors can be equivalently parametrized by the family of $4\pi$-periodic scalar functions $\phi(\varphi)$ such that
\begin{equation}
|\Phi(\varphi)\rangle = \frac{1}{\sqrt{2}} \begin{pmatrix} -i(\phi(\varphi)-\phi(\varphi+2\pi)) \\ \phi(\varphi)+\phi(\varphi+2\pi)\end{pmatrix}.
\end{equation} 
Then we readily check that Eq.~\eqref{eq:main} is equivalent to the scalar eigenproblem
\begin{equation}
\label{eq:main2}
\left(\sqrt{\Omega + \hat H}-\sqrt{2 E_J} \sin \frac{\hat \varphi}{2} \right)\psi(\varphi) =0,
\end{equation}
where
\begin{equation}
\label{eq:main3}
 \hat H =  \hat H_J + E_C(-2 i \partial_{\varphi} -{\cal N})^2-E_{{\rm min}}({\cal N}).
\end{equation}
Compared with Eq.~(4) of Ref.~\cite{Houzet2024}, we see that the chirality index $s$ has been assigned a given sign. Taking the opposite sign would correspond to considering the case where the Coulomb island is ``poisoned'' by an additional quasiparticle not taking part in the bound-state formation.

The method is easily extended to a setup formed of two islands on either side of the junction. In that case, $\varphi$ is the superconducting phase difference between the islands, and the charging energy depends on the difference of the charges in both islands.

\section{Bound-state energy}
\label{sec:3}

In this section we study the eigenspectrum defined by Eqs.~\eqref{eq:main2} and \eqref{eq:main3} in various regimes.

Let us consider first the regime, where the second term on the l.h.s. of Eq.~\eqref{eq:main2} is small. In a {Cooper-pair box,}  $E_J\leq E_J^*\ll E_C$,  the two lowest eigenstates $|0\rangle$ and $|1\rangle$ of $H$ have energies $E_{0,1}=E_C[({\cal N}-1/2)^2+1/4] \pm E_C( {\cal N}-1/2)$, such that $E_{\rm min}=\min(E_0,E_1)$. Here 0 and 1 label the number of excess charges in the island. The degeneracy of the states $|0\rangle$ and $|1\rangle$ at ${\cal N}=1/2$ is lifted by the  second term on the l.h.s.~of Eq.~\eqref{eq:main2}. Namely, projecting Eq.~\eqref{eq:main2} onto these states, we readily find that the device accommodates a single, spin-degenerate bound state with binding energy
\begin{equation}
\label{eq:ABS-charging}
\Omega=-E_C| {\cal N}-1/2|+\sqrt{E_C^2( {\cal N}-1/2)^2+E_J^2/4}
\end{equation}
if $|{\cal N}-1/2|\ll 1$, in agreement with Ref.~\cite{Pesin2004} (up to a missing factor 1/4 in front of $E_J$ in that reference). Note that $E_J^*$ does not appear in Eq.~\eqref{eq:ABS-charging} in the considered regime.

On the other hand, in {a transmon, $E_J^*\gg E_C$,} we may ignore the exponentially small gate modulation of eigenenergies, and approximate $H$ of Eq.~\eqref{eq:main3} as the Hamiltonian of a quantum harmonic oscillator with resonance frequency $\hbar\omega_0=\sqrt{8E^*_JE_C}$. Then $E_{\rm min}=-E_J^*+\hbar\omega_0/2$. 
Using linearization, $\sin\hat\varphi/2\approx\hat\varphi/2=(E_C/8E^*_J)^{1/4}(\hat b+\hat b^\dagger)$ with ladder operators $\hat b$ and $\hat b^\dagger$, and projecting Eq.~\eqref{eq:main2} onto the two lowest-energy states of the oscillator, we find perturbatively the binding energy
\begin{equation}
\label{eq:asymptote1}
\Omega=\hbar \omega_0/16N_{\rm ch}^2.
\end{equation} 
The correction remains small on the scale of $\hbar \omega_0$ if $N_{\rm ch}\gg 1$, 
{further requiring the condition $E_J^* \gg E_J$ for this result to hold.}

\begin{figure}
\centering
\includegraphics[width=.7\columnwidth]{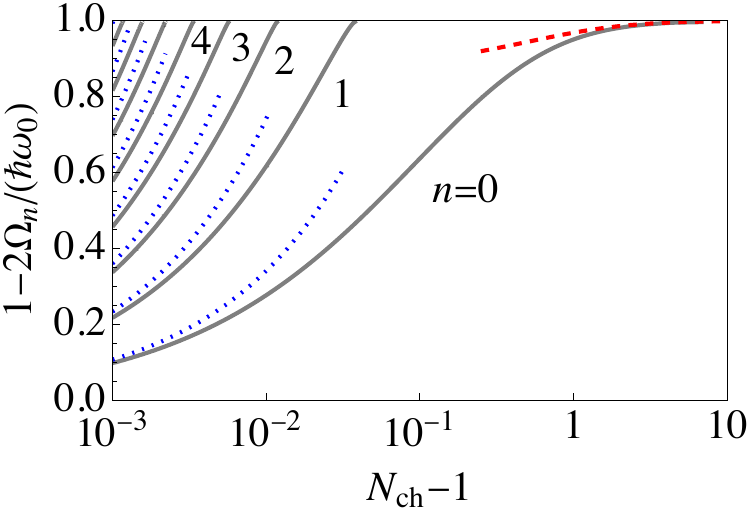}
\caption{\label{F1}
Bound state energies in the odd sector of a {transmon, $E^*_J\gg E_C$}. Each channel of the junction accommodates a finite number of bound states whose energy ${\cal E}_n=1-2\Omega_n/(\hbar \omega_0)$ is given as a function of $N_{\rm ch}=E_J^*/E_J$ (plain lines). The dashed and dotted lines give the asymptotes at $N_{\rm ch}\gg 1$ [Eq.~\eqref{eq:asymptote1}] and $0<N_{\rm ch}-1\ll 1$ [Eq.~\eqref{eq:asymptote2}], respectively.}
\end{figure}

As $N_{\rm ch}$ decreases, we may formulate the eigenproblem defined by Eq.~\eqref{eq:main2} in dimensionless units,
\begin{equation}
\label{eq:main-eq-v2}
\sqrt{\frac 2{\hbar\omega_0}}\hat {\cal H}_{\rm eff}(E)=\sqrt{\hat X^2+\hat P^2-{\cal E}}-\frac 1{\sqrt{N_{\rm{ch}}}}\hat X
\end{equation}
with $\hat X=(\hat b+\hat b^\dagger)/\sqrt{2}$, $\hat P=(\hat b-\hat b^\dagger)/i\sqrt{2}$, and ${\cal E}=2(E-\Delta+E_J^*)/\hbar\omega_0$. 
In particular, at $0<N_{\rm ch}-1\ll 1$, we anticipate that a zero-energy eigenstate of $\hat {\cal H}_{\rm eff}$ is localized in the region $X\gg 1$ (while still $X\ll (E_J^*/E_c)^{1/4}$ for the harmonic-oscillator expansion to be valid). Then we may approximate
\begin{eqnarray}
\label{eq:approximation}
\sqrt{\hat X^2+\hat P^2-{\cal E}}&=&\int \frac{d\zeta}{\pi}\frac{\hat X^2+\hat P^2-{\cal E}}{\hat X^2+\hat P^2-{\cal E}+\zeta^2}
\\
&\approx&\int \frac{d\zeta}{\pi}\frac{1}{\hat X^2+\zeta^2}\left[\hat X^2+(\hat P^2-{\cal E})\frac{\zeta^2}{\hat X^2+\zeta^2}\right]\nonumber\\
&=&\hat X+\frac 1{2\hat X}\left[\left(\hat P+\frac i{2 \hat X}\right)^2-{\cal E}+\frac 1{4\hat X^2}\right].\nonumber
\end{eqnarray}
Thus, finding an eigensolution of the equation $\hat {\cal H}_{\rm eff}(E)\Psi_E=0$ appears to be equivalent to finding an eigenfunction $\tilde \Psi_E(X)=\Psi_E(X)/\sqrt{X}$ of the effective (dimensionless) Hamiltonian
\begin{equation}
\label{eq:radial-H}
\hat {\cal H}=\hat P^2+(N_{\rm ch}-1)\hat X^2+\frac 1{4\hat X^2}
\end{equation}
defined on a semi-infinite line, $X>0$. We recognize Eq.~\eqref{eq:radial-H} as the radial harmonic oscillator Hamiltonian in three dimensions. Its eigenenergies are given as~\cite{Cooper1995} 
\begin{equation}
\label{eq:asymptote2}
{\cal E}_n=2\sqrt{N_{\rm ch}-1}(2n+1+1/\sqrt{2}),\qquad  n\in \mathbb{N}.
\end{equation}
The corresponding binding energies are given as $\Omega_n=\hbar\omega_0(1-{\cal E}_n)/2$. The spatial extension of the associated wavefunctions scales as $1/(N_{\rm ch}-1)^{1/4}$, ensuring the validity of the result if $E_C/E_J\ll N_{\rm ch}-1 \ll 1$. Furthermore, we estimate that approximately $1/\sqrt{N_{\rm ch}-1}$ bound states are accommodated below the continuum. We numerically solve the eigenproblem defined by Eq.~\eqref{eq:main-eq-v2} to find the $N_{\rm ch}$-dependence of the spectrum that interpolates between Eqs.~\eqref{eq:asymptote1} and \eqref{eq:asymptote2}, cf.~Fig.~\ref{F1}. 

As $N_{\rm ch}$ further decreases, such that $N_{\rm ch}-1\ll E_C/E_J$, the states localized near $\varphi=0$ and $2\pi$ overlap and the harmonic approximation is no longer suitable. To describe their hybridization, we use the same approximation as in Eq.~\eqref{eq:approximation} but keeping the sinusoidal terms. As a result, at $N_{\rm ch}=1$,  finding an eigensolution of the equation ${\cal H}_{\rm eff}(E)\Psi_E=0$ is equivalent to finding an eigenfunction $\tilde \Psi_E(\varphi)=\Psi_E(\varphi)/\sqrt{\sin(\varphi/2)}$  that solves the effective  Hamiltonian
\begin{equation}
\label{eq:Rosen-Morse}
\hat H=E_C \left[\hat N^2
+\frac {1}{4\tan^2 (\hat\varphi/2)}\right]
\end{equation}
with $\hat N=-2i\partial_\varphi$, defined in the interval $0<\varphi<2\pi$. Note that the relevant energy scale is $E_C\ll\hbar\omega_0$. We recognize the Hamiltonian for a particle in a Rosen-Morse potential, whose eigenenergies are given as~\cite{Cooper1995} 
\begin{equation}
\label{eq:asymptote3}
\tilde E_n={E_C}\left[\left(n +\frac{1 +\sqrt{2}}2\right)^2-\frac14\right], \qquad  n\in \mathbb{N}.
\end{equation}
The corresponding binding energies are given as $\Omega_n=\hbar\omega_0/2-\tilde E_n$.  
 
\begin{figure}
\includegraphics[width=0.45\columnwidth]{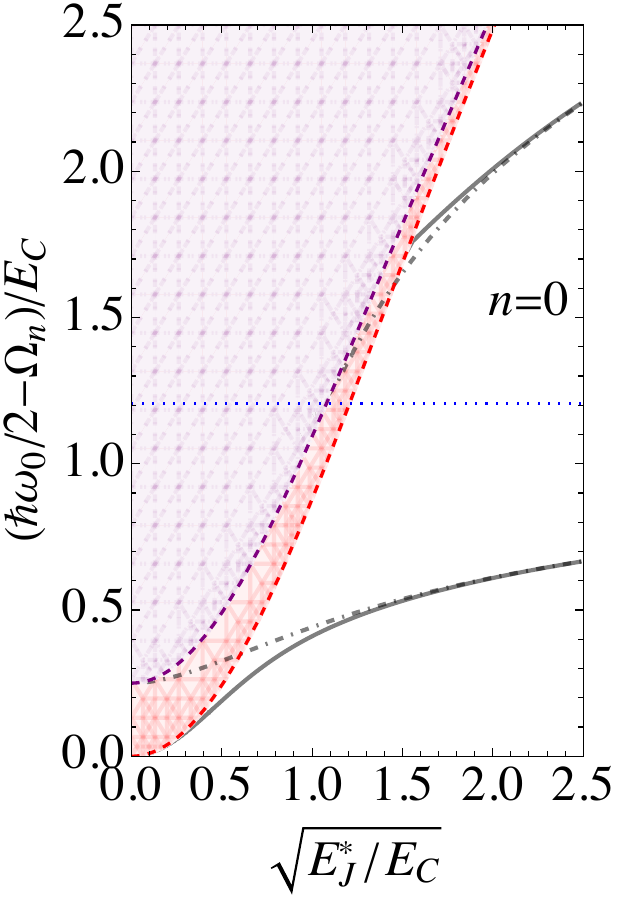}
\includegraphics[width=0.438\columnwidth]{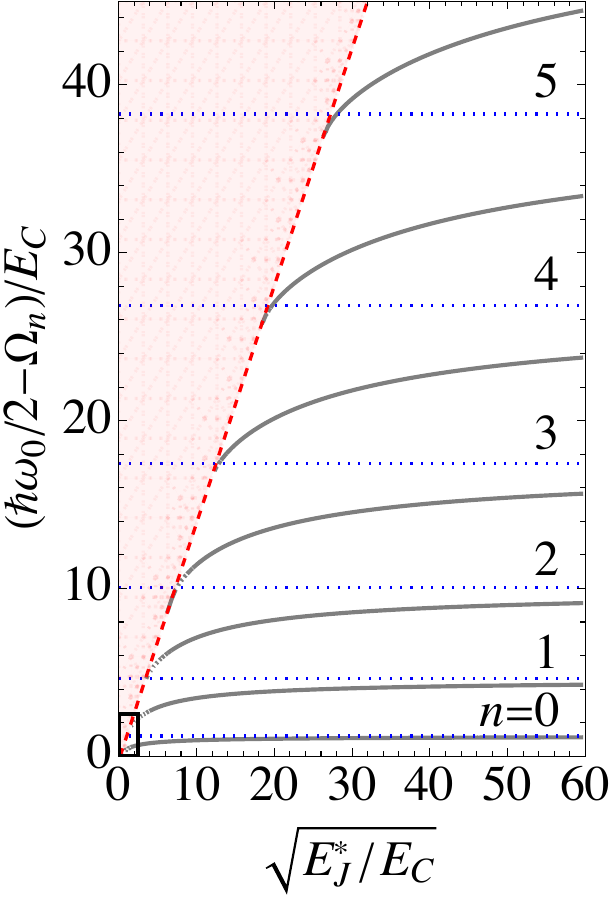}
\caption{\label{F2}
Bound state energies in the odd sector of a single-channel {superconducting qubit} as a function of $E_J^*/E_C$ at $N_{\rm ch}=1$ (plain line). 
The left panel zooms in the boxed region of the right panel to resolve the gate dependence of the spectrum between ${\cal N}=0$ (plain line) and ${\cal N}=1/2$ (dash-dotted line). The dotted lines give the asymptotes {in a transmon} at $E_J^*\gg E_C$ [Eq.~\eqref{eq:asymptote3}] and  the dashed lines give the edge of the continuum spectrum at ${\cal N}=0$ and ${\cal N}=1/2$, respectively.
}
\end{figure}

We have not been able to use the Wentzel-Kramers-Brillouin (WKB) method to explore the ${\cal N}$-modulation of the energies. We plot the spectrum obtained numerically at $N_{\rm ch}=1$ and for ${\cal N}=0,1/2$ in Fig.~\ref{F2}. 

\section{Conclusion}
\label{sec:4}

The realization of superconducting qubits relies on the coupling of a Josephson junction to its electromagnetic environment. While the most common superconducting qubits use the electromagnetic degrees of freedom only, superconducting quasiparticles play an important role -- either as a nuisance degrading qubit performance or as a feature enabling novel qubit architectures. These quasiparticles may be present in the leads or they may get trapped in Andreev bound states at the Josephson junction~\cite{Zgirski2011,Bretheau2013,Levenson2014}, such that the junction itself (rather than the Coulomb island, as discussed in Section~\ref{sec:1}) is \lq\lq poisoned\rq\rq. Our work addresses the physics of  a poisoned Josephson junction. While the effect of electron-electron interactions on the Andreev spectrum of a Josephson junction has been intensively studied, relatively few works have addressed the effect of the interaction with the electromagnetic environment of the Josephson junction. Building on our earlier work on a Josephson junction coupled to an ohmic environment, here we showed for the case of the Cooper-pair box and the transmon that this interaction leads to rich new physics, profoundly modifying the bound state spectrum. Our predictions could be probed by microwave spectroscopy~\cite{Bargerbos2020,Kringho2020} in forthcoming experiments with superconductor/semiconductor/superconductor junctions. 

The work opens several perspectives.  An obvious next step would be to extend our analysis to more complex qubit architectures involving additional circuit elements. Furthermore, while our work concentrated on the tunneling regime, Andreev bound state physics plays an even more important role at larger transmissions~\cite{Averin1999,Pikulin2019}. In that case, the first excited state in the even sector, corresponding to a doubly occupied Andreev bound state, is of particular interest as it may couple to the ground state on a much shorter time scale. Finally, spin effects could be taken into account allowing one to investigate Andreev spin qubits~\cite{Chtchelkatchev2003,Padurariu2010,Tosi2019,Hays2020,Hays2021,Pita-Vidal2023} that are operated in the odd parity sector.

\bmhead{Acknowledgements}

MH and JSM acknowledge funding from the Plan France 2030 through the project NISQ2LSQ ANR-22-PETQ-0006 and FERBO ANR-23-CE47-0004. YVN acknowledges support from the Université Grenoble Alpes for an extended stay in Grenoble during which most of the presented work was performed.

\end{document}